# Achievability of Quantum Regime in Optomechanics of Electromagnetically Levitated Nanoparticles: Inclusion of Gradient Force Fluctuations

Amir M. Jazayeri* and Khashayar Mehrany

*Department of Electrical Engineering, Sharif University of Technology, Tehran 145888-9694, Iran*

Counterintuitively, experiments show that an electromagnetically levitated particle escapes from its trap when the ambient pressure is reduced below a certain level even if the particle's motion is cooled by a resonator-based or feedback-based mechanism. Here, we theoretically show that the ambient pressure must be kept well above a critical value arising from gradient force fluctuations (viz., fluctuations in part of the EM force whose Hamiltonian is quadratic in the position of the particle). Also, we consider other force fluctuations, and determine whether different realizations of feedback cooling are able to reach the ground state. In some realizations, the cooling rate must be kept well below a critical value arising from feedback-induced gradient force fluctuations.

Although slow response of mechanical oscillators hinders the application of optomechanical phenomena [1], optomechanics offers unique opportunities in the quantum regime [2-4], which necessitate cooling the mechanical oscillators to near their ground state [5,6]. Usually, the mechanical mode to be cooled is a standing or travelling wave [5-7] within a material body whose center of mass is motionless. Recently, each component of the center-of-mass motion of an electromagnetically levitated particle around its trapping point has been proposed as an ideal mechanical oscillator [8-16] on the grounds that its damping rate can be made arbitrarily small simply by reducing the ambient pressure. However, due to a phenomenon, described as 'known yet unexplained' in [9], the particle escapes from the trap at small enough ambient pressures, whether its motion is cooled by means of a resonator [8-11] or feedback [12-16].

We analyze all possible realizations of feedback cooling proposed in the literature [12-16], and determine whether they are able to enter the quantum regime. Quantum fluctuations in the laser power ($P_L$) used to trap the particle (and illuminate it for photodetection) lead to gradient force fluctuations, radiation pressure fluctuations, and recoil force. We show that the particle escapes from the trap if the ambient pressure is not kept well above a critical value arising from the gradient force fluctuations. Our derived critical value is relevant in any other system employing an electromagnetically levitated particle, e.g. the resonator-based systems proposed in [8-11]. Quantum fluctuations in $P_L$ also lead to the measurement noise, and therefore to feedback-induced force fluctuations. We show that the particle escapes from the trap in some realizations of feedback cooling if the cooling rate is not kept well below a critical value arising from feedback-induced gradient force fluctuations.

*Thermal motion.–* We consider a small dielectric particle of mass $M$ levitated by the optical gradient force around the focal point of a lens whose axis is defined as the $z$ axis. The position (or position operator) of the particle center with respect to the focal point of the lens is denoted by $\vec{r} = (x_1, x_2, x_3)$. The gas molecules surrounding the particle exert a damping force $-M\Gamma\dot{\vec{r}}$ and a random force $(f_1, f_2, f_3)$ on the particle. The spectral density of $f_i$ (see [17] for the definition of spectral density), which is the same for all $i$, is proportional to the intrinsic damping rate $\Gamma$ [18]. This is a manifestation of the fluctuation-dissipation theorem, which relates the spectral density of $x_i$ to the imaginary part of its susceptibility. The variance of $x_i$ [19] can be written as $\mathrm{var}(x_i) = (2\bar{n}_{th,i} + 1)\hbar/(2M\Omega_i)$, where $\Omega_i$ denotes the oscillation frequency (determined by the optical trapping force), and $\bar{n}_{th,i}$ denotes the mean phonon number, which reads $k_B T/(\hbar\Omega_i)$ for $k_B T \gg \hbar\Omega_i$ in terms of the ambient temperature $T$. For the moment, we have ignored any fluctuations in the laser power.

The intrinsic damping rate ($\Gamma$) in a rarefied gas, which is proportional to the ambient pressure ($P_{am}$), is not given by Stokes' law, but rather by Epstein's formula [20]. Epstein's formula is applicable even if the surface temperature ($T_s$) of the particle is not equal to the ambient temperature. However, in such a case, $T$, which appeared in the expression of $\bar{n}_{th,i}$, denotes a temperature larger than the ambient temperature (for more details, see [21,22]).

It is noteworthy that the photophoretic force, which is a result of temperature gradient over the particle surface [23,24], is orders of magnitude smaller than the EM force in the system considered in this letter.

*EM force fluctuations.–* Under the dipole approximation, the classical EM force exerted on the particle can be written as the sum of the so-called gradient force $\vec{g} = (g_1, g_2, g_3)$ and radiation pressure $\vec{\rho} = (\rho_1, \rho_2, \rho_3)$ [25]. Around the focal point of the lens, $g_i$ acts as a spring force $-K_i x_i$, allowing us to define a mechanical mode with the oscillation frequency $\Omega_i = \sqrt{K_i / M}$, which appeared in the expression of $\bar{n}_{th,i}$. We write $K_i$ as $A_i P_L$, where $P_L$ denotes the power carried by the trapping beam, and $A_i$ is a coefficient given in [22]. We ignore $\rho_1$ and $\rho_2$ in that they can be approximated by $\kappa_1 x_1$ and $\kappa_2 x_2$, respectively, where $\kappa_1$ and $\kappa_2$ are *positive* and much smaller than $K_1$ and $K_2$. The component $\rho_3$ is almost insensitive to $\vec{r}$, and can be written as $B P_L$, where $B$ is a coefficient given in [22].

We now consider fluctuations $P_L - \bar{P}_L$ in the laser power, where $P_L$ now denotes the operator corresponding to the optical power carried by the trapping beam (and passing through the focal plane of the lens), and $\bar{P}_L$ denotes the expectation value of $P_L$. Assuming that the fluctuations $P_L - \bar{P}_L$ are solely due to inherent uncertainty in the emission of photons by the laser, the spectral density of $P_L$ [ $S_{P_L}(\omega)$ ] is the constant function $\hbar\omega_0 \bar{P}_L$.

The operator corresponding to $g_i$ reads $-A_i P_L x_i$. Insofar as the gradient force is concerned, the presence of the particle slightly changes the EM energy stored around the focal point of the lens, but does not change the number of the photons passing through the focal plane of the lens, hence the Hamiltonian $H_{g_i} = A_i P_L x_i^2 / 2$ commutes with $P_L$. The Hamiltonian is quadratic in $x_i$ because it has been written around the point where the gradient force vanishes.

One might infer from $\bar{\rho}_3 = B\bar{P}_L$ that the operator corresponding to $\rho_3$ reads $BP_L$. Such an inference is incorrect in that radiation pressure comes from the *initial* linear momentum of the photons interacting with (viz., absorbed or scattered by) the particle, and its Hamiltonian does not commute with $P_L$. The spectral density of $\rho_3$ can be written as $B'^2 S_{P_L}(\omega)$, where $B'$, which we derive in [22], is not equal to $B$. It should be noted that if the laser power fluctuations were mainly due to fluctuations in the electric current applied to the laser [viz., if $S_{P_L}(\omega)$, apart from a coefficient, was equal to the spectral density of the electric current applied to the laser], $B'$ would be equal to $B$. We will return to this point when we discuss feedback cooling.

The *final* linear momentum of the photons *scattered* by the particle begets a recoil force $\vec{\sigma}$, which is almost insensitive to $\vec{r}$. Unlike $g_i$ and $\rho_3$, the expectation value of $\sigma_i$ is zero. The spectral density of $\sigma_i$ can be written as $C_i^2 \hbar\omega_0 \bar{P}_L$, where $C_i$ is given in [22].

To find the mean phonon numbers, we adopt an approach which has also been used in conventional optomechanics [26-28]. By applying Fermi's golden rule [together with the assumption that the particle's motion does not significantly change $S_{P_L}(\omega)$] to the master equation, we can write the following equation for the phonon number $n_i(t)$:

$$\dot{P}_{i,m} = \Gamma_{g,\uparrow,i} m(m-1)P_{i,m-2} + (\bar{n}_{th,i}\Gamma + \Gamma_{r,\uparrow,i})mP_{i,m-1} - (\bar{n}_{th,i}\Gamma + \Gamma_{r,\uparrow,i})(m+1)P_{i,m} - [(\bar{n}_{th,i}+1)\Gamma + \Gamma_{r,\downarrow,i}]mP_{i,m}$$
$$-\Gamma_{g,\uparrow,i}(m+2)(m+1)P_{i,m} - \Gamma_{g,\downarrow,i}(m-1)mP_{i,m} + [(\bar{n}_{th,i}+1)\Gamma + \Gamma_{r,\downarrow,i}](m+1)P_{i,m+1} + \Gamma_{g,\downarrow,i}(m+1)(m+2)P_{i,m+2}$$
, (1)

where $P_{i,m}(t)$ is the probability that $n_i$ at $t$ equals $m$. The rate $\Gamma_{g,\uparrow/\downarrow,i}$ comes from $g_i$, and reads $S_{K_i}(\mp 2\Omega_i)/(4M\Omega_i)^2$, where $S_{K_i}(\omega)$ denotes the spectral density of $K_i = A_i P_L$. The rate $\Gamma_{r,\uparrow/\downarrow,i}$ comes from $\rho_i + \sigma_i$, and reads $[S_{\rho_i}(\mp\Omega_i) + S_{\sigma_i}(\mp\Omega_i)]/(2\hbar M\Omega_i)$, where $S_{\rho_i}(\omega)$ and $S_{\sigma_i}(\omega)$ denote the spectral densities of $\rho_i$ and $\sigma_i$, respectively. Since $\Gamma_{g,\uparrow,i}$ and $\Gamma_{r,\uparrow,i}$ are equal to $\Gamma_{g,\downarrow,i}$ and $\Gamma_{r,\downarrow,i}$, respectively, they will hereafter be denoted by $\Gamma_{g,i}$ and $\Gamma_{r,i}$, respectively.

The mean phonon number is $\bar{n}_i(t) = \sum_m m P_{i,m}(t)$. We are interested in steady state [viz., at large enough $t$, where $\dot{P}_{i,m}(t)$ all vanish, and $S_{x_i}(\omega)$ is definable]. By using Eq. (1), we find that the mean phonon number in steady state, which is shortened to 'mean phonon number' in this letter, reads $\bar{n}_i = (\bar{n}_{th,i}\Gamma + \Gamma_{r,i} + 4\Gamma_{g,i})/(\Gamma - 8\Gamma_{g,i})$.

Our derived expression of $\bar{n}_i$ indicates that $\Gamma$ must be kept well above the critical value $\Gamma_{cr} = \max_i(8\Gamma_{g,i})$ – otherwise, renormalization of the mechanical oscillation frequencies by the gradient force fluctuations becomes so strong that the mechanical modes disappear, and the particle escapes from the trap. Such a destruction of the mechanical modes is to some extent similar to the destruction of the Higgs mode [29] in the magnetically ordered phase of the quantum rotor model in low dimensions. Unlike the celebrated renormalization effects in conventional optomechanics [26-28], the renormalization effect here is wideband, in the sense that it affects the susceptibility and spectral density of $x_i$ in the frequency range from zero to $\Omega_i$. Therefore, $\Gamma$ must be kept well above $\Gamma_{cr}$ even in the presence of feedback cooling, which is narrow-band. It is noteworthy that the same argument applies to any other system employing electromagnetically levitated particles (e.g. the resonator-based systems proposed in [8-11]).

The mean phonon number is simplified to $\bar{n}_i = (\bar{n}_{th,i}\Gamma + \Gamma_{r,i})/\Gamma$ when $\Gamma \gg \Gamma_{cr}$. Interestingly, $\Gamma_{cr}$ is insensitive to $\bar{P}_L$ and the radius of the particle ($R$), but is an increasing function of the numerical aperture (*NA*) of the lens. The critical ambient pressure $P_{am,cr}$ corresponding to $\Gamma_{cr}$ is (almost) insensitive to $\bar{P}_L$, but is proportional to $R$, and is an increasing function of *NA*. As is usually the case in experiments [12-15], we assume that the beam trapping the particle illuminates it for photodetection as well – otherwise [16], the contribution of the illuminating beam to $\Gamma_{cr}$ and $\Gamma_{r,i}$ must be taken into account.

*Feedback cooling.–* The idea of feedback cooling is to measure $\vec{r}$, and exert a cooling force $\vec{F} = -M(\Gamma_{fb,1}\dot{x}_1, \Gamma_{fb,2}\dot{x}_2, \Gamma_{fb,3}\dot{x}_3)$ on the particle. In the case where $\vec{F}$ is an optical force, $F_i$ is accompanied by unwanted force components (UFC) of the form $\hat{j}\dot{x}_i$ (for each $j \neq i$) and of the same order of magnitude as $F_i$. We will see that UFC disables feedback cooling if conditions such as $|\Omega_i - \Omega_j| \gg \delta_i + \delta_j$ are not met, where $\delta_i \approx (\Gamma + \Gamma_{fb,i})/2 \approx \Gamma_{fb,i}/2$ is the linewidth of $S_{x_i}(\omega)$ in the presence of feedback cooling. In the case where $\vec{F}$ is a Coulomb force, UFC does not take place at all.

The other nonideality in feedback cooling comes from inherent uncertainty in the emission of photons by the laser illuminating (and trapping) the particle. The photodetector intended to measure $x_i$ generates a photocurrent $I_i$ whose spectral density, apart from an unimportant coefficient, can be written as $S_{x_i}(\omega) + S_{n_i}(\omega)$, where we have approximated $E[x_i(t)x_i(t+\tau)]$ by its real part, and assumed that the detection bandwidth ($W_i$) is large in comparison with $\Omega_i$. The unwanted $S_{n_i}(\omega)$, which we derive in [22], is a constant function of $\omega$. Since the axis of the beam illuminating (and trapping) the particle has been defined as the $z$ axis, $S_{n_3}$ is proportional to $d_{eff,3}^2 / a_{eff,3}$ while $S_{n_1}$ and $S_{n_2}$ are proportional to $(d_{eff,1}^2 / a_{eff,1})^2$ and $(d_{eff,2}^2 / a_{eff,2})^2$, respectively, where $d_{eff,i}$ and $a_{eff,i}$ are parameters we name the 'effective distance' and 'effective area' of the photodetector $i$. The ratio $d_{eff,i}^2 / a_{eff,i}$ cannot be made smaller than a lower bound which is approximately equal to $50\pi$ for $i=3$, and $450\pi$ for $i=1,2$. It is noteworthy that $\mathrm{var}(n_i) \approx 2W_i S_{n_i}$ may be much larger than $\mathrm{var}(x_i)$. However, the variance of $n_i$ as seen by the particle (viz., $4\delta_i S_{n_i}$) is usually much smaller than $\mathrm{var}(x_i)$ except when the mechanical mode is cooled to near its ground state.

In the case where $\vec{F}$ is an optical force, the mean power ($\bar{P}_i$) of the light intended to generate $F_i$ is much smaller than $\bar{P}_L$, and we can therefore assume that $\vec{F}$ is not accompanied by a recoil force. Also, unlike the fluctuations $P_L - \bar{P}_L$, which are due to inherent uncertainty in the emission of photons, the fluctuations $P_i - \bar{P}_i$ are mainly due to fluctuations in the electric current applied to the laser. More precisely, $S_{P_i}(\omega)$, apart from a coefficient, reads $S_{x_i}(\omega) + S_{n_i}(\omega)$. The gradient force fluctuations due to $n_i$ lead to a rate $\tilde{\Gamma}_{g,i}$ similar to $\Gamma_{g,i}$, and can destroy the mechanical modes. To avoid the destruction of the mechanical modes, one must keep $\Gamma_{fb,i}$ well below a critical value $\Gamma_{fb,cr,i}$ (sometimes, one can filter the electric currents instead). The radiation pressure fluctuations due to $n_i$ lead to a rate $\tilde{\Gamma}_{r,i}$ similar to $\Gamma_{r,i}$. Assuming that $\Gamma_{cr} \ll \Gamma$ and $\Gamma \ll \Gamma_{fb,i} \ll \Gamma_{fb,cr,i}$, the mean phonon number reads

$$\bar{n}_i = (\bar{n}_{th,i}\Gamma + \Gamma_{r,i} + \tilde{\Gamma}_{r,i})/(\Gamma + \Gamma_{fb,i}) \simeq (\bar{n}_{th,i}\Gamma + \Gamma_{r,i} + \tilde{\Gamma}_{r,i})/\Gamma_{fb,i}. \qquad (2)$$

We will derive $\Gamma_{fb,cr,i}$ and $\tilde{\Gamma}_{r,i}$ in all possible realizations of feedback cooling. In the case where $\vec{F}$ is a Coulomb force, the Coulomb force fluctuations accompanying $\vec{F}$ lead to a rate $\tilde{\Gamma}_{c,i}$ (in place of $\tilde{\Gamma}_{r,i}$), but $\tilde{\Gamma}_{g,i}$ is zero (viz., $\Gamma_{fb,cr,i}$ approaches infinity).

*Feedback cooling by radiation pressure.–* In this realization, the components of $\vec{F}$ are generated by three lasers, which are all distinct from the laser trapping (and illuminating) the particle [12]. The laser responsible for generating $F_j = -M\Gamma_{fb,j}\dot{x}_j$ creates a beam with the optical power $P_j = \bar{P}_j - M\Gamma_{fb,j}(\dot{x}_j + \dot{n}_j)/\tilde{B}_{jj}$, where $\bar{P}_j \ll \bar{P}_L$. This beam exerts a radiation pressure $\hat{j}(F_j + \tilde{\rho}_{n,j})$ and a gradient force $\hat{i}(\tilde{g}_{ji} + \tilde{g}_{n,ji})$ (for each $i$) on the particle, where $\tilde{\rho}_{n,j}$, $\tilde{g}_{ji}$, and $\tilde{g}_{n,ji}$ read $-M\Gamma_{fb,j}\dot{n}_j$, $(\tilde{A}_{ji}/\tilde{B}_{jj})M\Gamma_{fb,j}\dot{x}_j x_i$, and $(\tilde{A}_{ji}/\tilde{B}_{jj})M\Gamma_{fb,j}\dot{n}_j x_i$, respectively. The coefficients $\tilde{B}_{jj}$ and $\tilde{A}_{ji}$ are defined for the cooling beam $j$ in the same way as $B$ and $A_i$ were defined for the trapping beam. For the sake of simplicity, we assume that the three cooling beams have identical wavelength, and identical beam radius. Therefore, $\tilde{B}_{jj}^2$ and $\sum_i \tilde{A}_{ij}^2$ (which is equal to $\sum_i \tilde{A}_{ji}^2$) are independent of $j$. We define $\tilde{a}_{ji} \triangleq \tilde{A}_{ji}/\tilde{B}_{jj}$ and $\tilde{a}^2 \triangleq \sum_j \tilde{a}_{ji}^2$.

The component $\tilde{g}_{ji}$ does not affect $\text{var}(x_i)$, provided that $|\Omega_j - 2\Omega_i| \gg \delta_j + 2\delta_i$. Otherwise, UFC (viz., the presence of $\tilde{g}_{ji}$ for $j \neq i$) disables feedback cooling. The rate $\tilde{\Gamma}_{r,i}$ in this realization of feedback cooling comes from $\tilde{\rho}_{n,i}$, and reads $M\Omega_i\Gamma_{fb,i}^2 S_{n_i}/(2\hbar)$, where

$S_{\dot{n}_i}(\omega)$ has been approximated by $\omega^2 S_{n_i}$. Also, the rate $\tilde{\Gamma}_{g,i}$ comes from $\sum_j \tilde{g}_{n,ji}$, and reads $\sum_j \tilde{a}_{ji}^2 \Gamma_{fb,j}^2 S_{n_j} / 4$. To avoid the destruction of the mechanical modes, $8\tilde{\Gamma}_{g,i}$ (for all $i$) must be kept well below $\Gamma$, or, equivalently, $\Gamma_{fb,i}^2$ must be kept well below $\Gamma_{fb,cr,i}^2 = \Gamma/(2\tilde{a}^2 S_{n_i})$.

One may filter the photocurrent $I_j$ (or the current $\dot{I}_j$) for each $j$ in a way that the information about $x_j$ remains intact while $n_j$ (or $\dot{n}_j$) converts into fluctuations whose spectral density is localized around $\pm \Omega_j$. In such a case, $\tilde{g}_{n,ji}$ (for each $i$) convert into force fluctuations which do not affect var($x_i$). As a result, $\tilde{\Gamma}_{g,i}$ becomes zero (viz., $\Gamma_{fb,cr,i}$ approaches infinity). However, $\tilde{\Gamma}_{r,i}$ remains intact.

The mean phonon number ($\bar{n}_i$) in Eq. (2) is minimized when $\Gamma_{fb,i}$ is equal to $\Gamma_{fb,opt,i} \simeq \sqrt{(\bar{n}_{th,i}\Gamma + \Gamma_{r,i})2\hbar/(M\Omega_i S_{n_i})}$, which is usually larger than $\Gamma_{fb,cr,i}$. The resulting mean phonon number ($\bar{n}_{min,i}$), which is equal to $2\sqrt{(\bar{n}_{th,i}\Gamma + \Gamma_{r,i})M\Omega_i S_{n_i}/(2\hbar)}$, is a decreasing function of $\bar{P}_L$ in that $\bar{n}_{th,i}\Omega_i S_{n_i} \propto T S_{n_i}$ is a decreasing function of $\bar{P}_L$ while $\Gamma_{r,i}\Omega_i S_{n_i} \propto \bar{P}_L S_{n_i}$ is independent of $\bar{P}_L$. Also, $\bar{n}_{min,i}$ is a decreasing function of $R$ in that $\bar{n}_{th,i}M\Omega_i S_{n_i} \propto TM S_{n_i}$ is a decreasing function of $R$ while $\Gamma_{r,i}M\Omega_i S_{n_i} \propto (B_i'^2 + C_i^2)S_{n_i}$ is independent of $R$. For a fixed $\Gamma$, $\bar{n}_{min,i}$ is a decreasing function of $NA$ in that $\bar{n}_{th,i}\Omega_i S_{n_i} \propto T S_{n_i}$ is a decreasing function of $NA$ while $\Gamma_{r,i}\Omega_i S_{n_i} \propto (B_i'^2 + C_i^2)S_{n_i}$ is independent of or a decreasing function of $NA$. However, for $\Gamma \approx \Gamma_{cr}$, $\bar{n}_{min,1}$ and $\bar{n}_{min,2}$ become increasing functions of $NA$ in that $\Gamma_{cr}$ is an increasing function of $NA$. It is emphasized that $\Gamma_{cr}$ is insensitive to $\bar{P}_L$ and $R$. When choosing $\bar{P}_L$, $R$, and $NA$, it should be noted that the surface temperature of the particle ($T_s$) must be kept below the melting point.

It is noteworthy that if the circuit differentiating the photocurrent $I_j$ generates fluctuations in excess of $\dot{n}_j$, they can be prevented from contributing to $\Gamma_{fb,cr,i}$ (for all $i$) by filtering the current $\dot{I}_j$, but they contribute to $\tilde{\Gamma}_{r,j}$ anyway.

*Feedback cooling by gradient force.–* In this realization, the laser trapping (and illuminating) the particle cools its motion as well [14-16]. Since the gradient force is linear in $x_i$, the current $\dot{I}_i$ must be multiplied by the current $I_i$ before being applied to the laser. Strictly speaking, the resulting cooling force component ($i\hat{g}_{ii} = -i\hat{\eta}_i x_i^2 \dot{x}_i$) is not of the form $-\hat{i}M\Gamma_{fb,i}\dot{x}_i$. However, we write $\eta_i$ as $M\Gamma_{fb,i}/\overline{x_i^2}$, and approximate $\hat{g}_{ii}$ by $-M\Gamma_{fb,i}\dot{x}_i$, where, in general, $\overline{x_i^2} = \mathrm{E}[x_i^2]$ must be calculated self-consistently.

The cooling force component $\hat{g}_{jj}$ is accompanied by a gradient force $i\hat{g}_{ji}$ (for each $i \neq j$), a gradient force $i\hat{g}_{n,ji}$ (for each $i$), and a radiation pressure $\hat{z}(\hat{\rho}_{j3} + \hat{\rho}_{n,j3})$, where, $\hat{g}_{ji}$, $\hat{g}_{n,ji}$, $\hat{\rho}_{j3}$, and $\hat{\rho}_{n,j3}$ read $-(A_i/A_j)\eta_j \dot{x}_j x_j x_i$, $-(A_i/A_j)\eta_j(\dot{x}_j n_j + x_j \dot{n}_j)x_i$, $(B/A_j)\eta_j \dot{x}_j x_j$, and $(B/A_j)\eta_j(\dot{x}_j n_j + x_j \dot{n}_j)$, respectively, and we have ignored $\dot{n}_j n_j$. We define $a_{ji} \triangleq A_i/A_j$ and $b_j \triangleq B/A_j$.

The component $\hat{g}_{ji}$ (for $i \neq j$) does not affect $\mathrm{var}(x_i)$, provided that $|\Omega_j - \Omega_i| \gg \delta_j + \delta_i$. The component $\hat{\rho}_{j3}$ does not affect $\mathrm{var}(x_3)$, provided that $|2\Omega_j - \Omega_3| \gg 2\delta_j + \delta_3$. Otherwise, UFC (viz., the presence of $\hat{g}_{ji}$ for $i \neq j$, and the presence of $\hat{\rho}_{j3}$) disables feedback cooling. The rates $\tilde{\Gamma}_{r,1}$ and $\tilde{\Gamma}_{r,2}$ in this realization of feedback cooling are negligible. However, the rate $\tilde{\Gamma}_{r,3}$ comes from $\sum_j \hat{\rho}_{n,j3}$, and reads $M\Omega_3 \sum_j b_j^2 \Gamma_{fb,j}^2 S_{n_j}/(2\hbar \overline{x_j^2})$, where $S_{\dot{n}_j x_j + n_j \dot{x}_j}(\omega)$ has been

approximated by $\omega^2 \overline{x_j^2} S_{n_j}$. Also, the rate $\tilde{\Gamma}_{g,i}$ comes from $\sum_j \hat{g}_{n,ji}$, and is found to be $\sum_j a_{ji}^2 \Gamma_{fb,j}^2 S_{n_j}/(4\overline{x_j^2})$. To avoid the destruction of the mechanical modes, $8\tilde{\Gamma}_{g,i}$ (for all $i$) must be kept well below $\Gamma$.

One may filter the photocurrent $I_j$ (or the current $I_j \dot{I}_j$) for each $j$ in a way that the information about $x_j$ remains intact while $n_j$ (or $\dot{x}_j n_j + x_j \dot{n}_j$) converts into fluctuations whose spectral density is localized around $\pm \Omega_j$ (or $\pm 2\Omega_j$). In such a case, $\hat{\rho}_{n,j3}$ and $\hat{g}_{n,ji}$ (for all $i \neq j$) convert into force fluctuations which do not affect $\text{var}(x_3)$ and $\text{var}(x_i)$, respectively. As a result, $\tilde{\Gamma}_{r,3}$ becomes zero. Also, $\tilde{\Gamma}_{g,i}$ decreases and becomes equal to $\Gamma_{fb,i}^2 S_{n_i}/(4\overline{x_i^2})$. Therefore, the condition that $8\tilde{\Gamma}_{g,i}$ (for all $i$) must be kept well below $\Gamma$ becomes equivalent to requiring that $\Gamma_{fb,i}^3$ is well below $\Gamma_{fb,cr,i}^3$, which is found to be $(\overline{n}_{th,i}\Gamma + \Gamma_{r,i})\Gamma \hbar/(2M\Omega_i S_{n_i})$, where we have replaced $\overline{x_i^2}$ by $\text{var}(x_i) \simeq 2\overline{n}_i \hbar/(2M\Omega_i)$, and $\overline{n}_i$ by $(\overline{n}_{th,i}\Gamma + \Gamma_{r,i})/\Gamma_{fb,i}$.

It is noteworthy that if the circuit differentiating the photocurrent $I_j$ generates fluctuations in excess of $\dot{n}_j$ (or if the circuit multiplying $\dot{I}_j$ by $I_j$ generates fluctuations in excess of $\dot{x}_j n_j + x_j \dot{n}_j$), they can be prevented from contributing to $\tilde{\Gamma}_{3,r}$ and $\Gamma_{fb,cr,i}$ (for all $i \neq j$) by filtering the current $\dot{I}_j I_j$, but they contribute to $\Gamma_{fb,cr,j}$ anyway.

*Feedback cooling by Coulomb force.–* In this realization, the particle has a net electric charge, and the cooling force components are generated by three capacitors [13]. The capacitor responsible for generating $F_i = -M\Gamma_{fb,i} \dot{x}_i$ creates a quasi-static electric field which exerts a Coulomb force $\hat{i}(F_i - M\Gamma_{fb,i} \dot{n}_i)$ on the particle. In this realization, UFC is negligible. Also, $\tilde{\Gamma}_{g,i}$ is zero, or, equivalently, $\Gamma_{fb,cr,i}$ approaches infinity. However, a rate $\tilde{\Gamma}_{c,i}$ (in place of $\tilde{\Gamma}_{r,i}$), which comes from $-\hat{i}M\Gamma_{fb,i}\dot{n}_i$, reads $M\Omega_i\Gamma_{fb,i}^2 S_{n_i}/(2\hbar)$. Therefore, $\bar{n}_i$ in Eq. (2) is minimized when $\Gamma_{fb,i}$ is equal to an optimum value $\Gamma_{fb,opt,i}$. The expressions of $\Gamma_{fb,opt,i}$ and the resulting $\bar{n}_{min,i}$ are the same as the expressions given above for feedback cooling by radiation pressure.

*Numerical examples.*– We consider a particle of fused silica and radius $R$. We assume that the parameters of the laser and the lens employed to generate the Gaussian beam trapping and illuminating the particle are $\lambda_0 = 1064$ nm, $\bar{P}_L = 100$ mW, and $NA=0.8$, which are the same parameters as in [14]. The relative permittivity of fused silica at $\lambda_0$ is 2.1+j10$^{-5}$.

For $R=70$ nm, the calculated oscillation frequencies $\Omega_1$ (and $\Omega_2$) and $\Omega_3$ are $2\pi\times367$ KHz and $2\pi\times208$ KHz, respectively. Due to UFC, we must assume that $\Omega_1$ and $\Omega_2$ are not exactly equal (viz., the lens must not be exactly symmetrical) in one realization of feedback cooling (viz., in feedback cooling by gradient force).

The calculated critical values $\Gamma_{cr}$ and $P_{am,cr}$ are $2\pi\times791$ nHz and $7\times10^{-10}$ mbar, respectively. This is in agreement with the ambient pressures reported in [15]. The calculated temperatures $T_s$ and $T$ are 1467 K and 697 K, respectively, when $P_{am}=7\times10^{-9}$ mbar. For a larger particle with $R=180$ nm, the oscillation frequencies and $\Gamma_{cr}$ remain unchanged, but $P_{am,cr}$, $T_s$, and $T$ are $2\times10^{-9}$ mbar, 1857 K (at $P_{am}=2\times10^{-8}$ mbar), and 866 K (at $P_{am}=2\times10^{-8}$ mbar), respectively.

For photodetectors, we assume that $a_{eff,1}$ and $a_{eff,3}$ are equal to $\lambda_0 d_{eff}/(45\pi)$ and $\lambda_0 d_{eff}/(5\pi)$, respectively (viz., their maximum allowable values). Also, we assume that $d_{eff,1}$ and $d_{eff,3}$ are equal to $d_{eff}=10\lambda_0$ (viz., their minimum allowable values). We first investigate feedback cooling by radiation pressure. We assume that the wavelength of the lasers and the numerical aperture of the lenses employed to generate the Gaussian beams cooling the particle's motion are 532 nm and 0.8. For the case with $R=70$ nm and $P_{am}=7\times10^{-9}$ mbar, the calculated critical values $\Gamma^2_{fb,cr,1}$ and $\Gamma^2_{fb,cr,3}$ are $(2\pi\times520$ mHz$)^2$ and $(2\pi\times13$ Hz$)^2$, respectively. The calculated mean phonon numbers are $\bar{n}_1=2\times10^3$ and $\bar{n}_3=1\times10^2$ when $\Gamma^2_{fb,i}=0.1\Gamma^2_{fb,cr,i}$. The mean phonon numbers can be reduced to $\bar{n}_1=3\times10^2$ and $\bar{n}_3=12$ by filtering the electric currents, and setting $\Gamma_{fb,1}=\Gamma_{fb,opt,1}=2\pi\times2$ Hz and $\Gamma_{fb,3}=\Gamma_{fb,opt,3}=2\pi\times94$ Hz.

For the case with $R=180$ nm and $P_{am}=2\times10^{-8}$ mbar, even if we increase $d_{eff}$ to $100\lambda_0$, and even if we do not filter the electric currents, the mean phonon numbers are $\bar{n}_1=2\times10^3$ and $\bar{n}_3$ =16 when $\Gamma^2_{fb,i}=0.1\Gamma^2_{fb,cr,i}$. The critical values $\Gamma^2_{fb,cr,1}$ and $\Gamma^2_{fb,cr,3}$ are $(2\pi\times15$ Hz$)^2$ and $(2\pi\times1.2$ KHz$)^2$, respectively. Our numerical results suggest that feedback cooling by radiation pressure is able to cool the $z$ component of the particle's motion to near its ground state. However, it requires $d_{eff}$ to be very small (viz., as small as $10\lambda_0$-$100\lambda_0$). The $z$ component has an advantage over the other two components in that the axis of the beam illuminating (and trapping) the particle is the $z$ axis, and therefore, $S_{n_3}$ is much smaller than $S_{n_1}$ and $S_{n_2}$.

Feedback cooling by Coulomb force is similar to feedback cooling by radiation pressure except that $\Gamma_{fb,cr,i}$ (for all $i$) approaches infinity for the former. In other words, one can always set $\Gamma_{fb,i}=\Gamma_{fb,opt,i}$ in feedback cooling by Coulomb force without filtering the electric currents.

We now investigate feedback cooling by gradient force. We only consider the case with $R=180$ nm and $P_{am}=2\times10^{-8}$ mbar. We assume that $d_{eff}$ is equal to $10\lambda_0$. Also, we assume that the electric currents are filtered. The calculated critical values $\Gamma^3_{fb,cr,1}$ and $\Gamma^3_{fb,cr,3}$ are $(2\pi\times58$

mHz)$^3$ and $(2\pi\times741$ Hz)$^3$, respectively. The calculated mean phonon numbers are $\bar{n}_1=1.5\times10^4$ and $\bar{n}_3=2\times10^3$ when $\Gamma^3_{fb,i}=0.1\Gamma^3_{fb,cr,i}$. Our numerical results suggest that feedback cooling by gradient force is unable to cool any component of the particle's motion to near its ground state.

* jazayeri@ee.sharif.edu

1. D. B. Sohn, S. Kim, and G. Bahl, Nat. Phot. 12, 91 (2018).
2. M. J. Hartmann and M. B. Plenio, Phys. Rev. Lett. 101, 200503 (2008).
3. Y. D. Wang and A. A. Clerk, Phys. Rev. Lett. 108, 153603 (2012).
4. T. A. Palomaki, J. W. Harlow, J. D. Teufel, R. W. Simmonds, and K. W. Lehnert, Nature 495, 210 (2013).
5. J. D. Teufel, T. Donner, D. Li, J. W. Harlow, M. S. Allman, K. Cicak, A. J. Sirois, J. D. Whittaker, K. W. Lehnert, and R. W. Simmonds, Nature 475, 359 (2011).
6. J. Chan, T. P. M. Alegre, A. H. Safavi-Naeini, J. T. Hill, A. Krause, S. Gröblacher, M. Aspelmeyer, and O. Painter, Nature 478, 89 (2011).
7. G. Bahl, M. Tomes, F. Marquardt, and T. Carmon, Nat. Phys. 8, 203 (2012).
8. D. E. Chang, C. A. Regal, S. B. Papp, D. J. Wilson, J. Ye, O. Painter, H. J. Kimble, and P. Zoller, Proc. Natl. Acad. Sci. U.S.A. 107, 1005 (2010).
9. N. Kiesel, F. Blaser, U. Delić, D. Grass, R. Kaltenbaek, and M. Aspelmeyer, Proc. Natl. Acad. Sci. U.S.A. 110, 14180 (2013).
10. U. Delić, M. Reisenbauer, D. Grass, N. Kiesel, V. Vuletić, and M. Aspelmeyer, Phys. Rev. Lett. 122, 123602 (2019).
11. D. Windey, C. Gonzalez-Ballestero, P. Maurer, L. Novotny, O. Romero-Isart, and R. Reimann, Phys. Rev. Lett. 122, 123601 (2019).
12. T. Li, S. Kheifets and M. G. Raizen, Nat. Phys. 7, 527 (2011).
13. F. Tebbenjohanns, M. Frimmer, A. Militaru, V. Jain, and L. Novotny, arXiv:1812.09875.
14. J. Gieseler, B. Deutsch, R. Quidant, and L. Novotny, Phys. Rev. Lett. 109, 103603 (2012).
15. V. Jain, J. Gieseler, C. Moritz, C. Dellago, R. Quidant, and L. Novotny, Phys. Rev. Lett. 116, 243601 (2016).
16. B. Rodenburg, L. P. Neukirch, A. N. Vamivakas, and M. Bhattacharya, Optica 3, 318 (2016).
17. We denote the spectral density of an observable $O$ by $S_O(\omega)$, and define it as the Fourier transform of $R_O(\tau)=\mathrm{E}[O(t)O(t+\tau)]-\mathrm{E}^2[O(t)]$ with respect to $\tau$, provided that the expectation values, denoted by E[·], are independent of $t$. The spectral density is a real function, and sometimes an even function. It is always an even function in the classical limit.
18. A. O. Caldeira and A. J. Leggett, Phys. Rev. Lett. 46, 211 (1981).
19. We denote the variance of an observable $O$ by var$[O]$, and define it as $\mathrm{E}[O^2(t)]-\mathrm{E}^2[O(t)]$, provided that the expectation values, denoted by E[·], are independent of $t$.
20. P. S. Epstein, Phys. Rev. 23, 710 (1924): Equation (29).
21. J. Millen, T. Deesuwan, P. Barker, and J. Anders, Nat. Nanotechnol. 9, 425 (2014).
22. See Supplemental Material at ?.
23. D. W. Keith, Proc. Natl. Acad. Sci. U.S.A. 107, 16428 (2010).
24. V. G. Shvedov, A. V. Rode, Y. V. Izdebskaya, A. S. Desyatnikov, W. Krolikowski, and Y. S. Kivshar, Phys. Rev. Lett. 105, 118103 (2010).
25. P. C. Chaumet and A. Rahmani, Opt. Exp. 17, 2224 (2009).

26. I. Wilson-Rae, N. Nooshi, W. Zwerger, and T. J. Kippenberg, Phys. Rev. Lett. 99, 093901 (2007).
27. F. Marquardt, J. P. Chen, A. A. Clerk, and S. M. Girvin, Phys. Rev. Lett. 99, 093902 (2007).
28. A. Nunnenkamp, K. Børkje, J. G. E. Harris, and S. M. Girvin, Phys. Rev. A 82, 021806 (2010).
29. S. Sachdev, *Quantum Phase Transitions* (Cambridge University Press, New York, 2011).

Supplemental Material

# Achievability of Quantum Regime in Optomechanics of Electromagnetically Levitated Nanoparticles: Inclusion of Gradient Force Fluctuations

Amir M. Jazayeri* and Khashayar Mehrany

*Department of Electrical Engineering, Sharif University of Technology, Tehran 145888-9694, Iran*

## A. Classical EM force

The electric field of the Gaussian beam trapping the particle (and illuminating it for photodetection) reads

$$\vec{E}_L = \hat{x} \operatorname{Re}(e^{-i\omega_0 t} E_{inc}) = \hat{x} \operatorname{Re}[e^{-i\omega_0 t} \frac{E_0}{\sqrt{1+X_3^2/z_0^2}} e^{-\frac{X_1^2+X_2^2}{w_0^2(1+X_3^2/z_0^2)} + ik_0 X_3 + i\frac{k_0(X_1^2+X_2^2)}{2X_3(1+z_0^2/X_3^2)} - i\arctan(X_3/z_0)}],$$

(A1)

where $(X_1, X_2, X_3)$ is the position of the observation point with respect to the focal point of the lens employed to generate the Gaussian beam, $z_0 = 2/(k_0 NA^2)$ is the Rayleigh range, $w_0 = 2/(k_0 NA)$ is the minimum beam radius, $NA$ denotes the numerical aperture of the lens, $k_0 = 2\pi/\lambda_0 = \omega_0/c$ is the wavenumber, and $c$ denotes the speed of light in free space [1]. The power carried by the Gaussian beam can be written as $P_L = \pi w_0^2 E_0^2 / (4\eta_0)$, where $\eta_0$ is the impedance of free space.

The dipole approximation assumes that the EM fields radiated by the particle, whose size is small in comparison with $\lambda_0$, are almost equal to the EM fields radiated by a point-like dipole in free space [2,3]. By applying the dipole approximation to the Maxwell stress tensor [4], the classical EM force exerted by the Gaussian beam on the particle is simplified to $\vec{F} = 0.5\,\text{Re}(\alpha E_{inc} \nabla E_{inc}^*)$, where $\alpha$ denotes the polarizability of the particle, $E_{inc}$ is given by Eq. (A1), and $E_{inc}$ and $\nabla E_{inc}^*$ are both evaluated at the position of the particle center [5]. The classical EM force can be rewritten as $\vec{F} = \vec{g} + \vec{\rho}$, where $\vec{g} = 0.25\alpha_R \nabla(|E_{inc}|^2)$ and $\vec{\rho} = 0.5\alpha_I \,\text{Im}[E_{inc}^* \nabla(E_{inc})]$ are the so-called gradient force and radiation pressure, respectively, and $\alpha_R$ and $\alpha_I$ denote the real and imaginary parts of $\alpha$. Assuming that the particle is a sphere of radius $R$ and relative permittivity $\varepsilon$, its polarizability reads $\alpha = \alpha_0 / [1 - ik_0^3 \alpha_0 / (6\pi\varepsilon_0)]$, where $\alpha_0$ denotes $4\pi\varepsilon_0 R^3 (\varepsilon - 1)/(\varepsilon + 2)$ [2,3]. If the particle has low loss [viz., when $\text{Im}(\varepsilon) \ll \text{Re}(\varepsilon)$], $\alpha_R$ and $\alpha_I$ can be approximated by $4\pi\varepsilon_0 a R^3$ and $8\pi\varepsilon_0 a^2 k_0^3 R^6 / 3$, respectively, where $a$ denotes $(\varepsilon_R - 1)/(\varepsilon_R + 2)$, and $\varepsilon_R$ denotes $\text{Re}(\varepsilon)$. It is noteworthy that $\vec{g}$ can also be derived by applying the dipole approximation to the method of virtual work rather than to the Maxwell stress tensor [6].

Since the particle is around the focal point of the lens (viz., $|\vec{r}| \ll \lambda_0$), the calculated $\vec{g}$ can be approximated by a spring force $-(K_1 x_1, K_2 x_2, K_3 x_3)$, where $\vec{r} = (x_1, x_2, x_3)$ is the position of the particle center. The stiffness vector $(K_1, K_2, K_3)$ can be written as $(A_1, A_2, A_3)P_L$, where $A_3$ and $A_1 = A_2$ are found to be $aNA^6 k_0^4 R^3 / (2c)$ and $aNA^4 k_0^4 R^3 / c$, respectively. Since the Gaussian beam in Eq. (A1) is symmetrical, $A_1$ and $A_2$ are equal. However, as is discussed in the main text, what we name UFC disables one realization of feedback cooling (viz., feedback cooling by gradient force) if the oscillation frequencies $\Omega_i = \sqrt{K_i / M}$ and $\Omega_j = \sqrt{K_j / M}$ (for any $j \neq i$) are equal. Therefore, the trapping beam must not be exactly symmetrical in that realization of feedback cooling.

Around the focal point of the lens, the calculated $\rho_3$ is almost insensitive to $\vec{r}$, and can be written as $BP_L$, where $B$ is found to be $4a^2(1-0.5NA^2)NA^2k_0^6R^6/(3c)$. The components $\rho_1$ and $\rho_2$ can be approximated by $\kappa_1 x_1$ and $\kappa_2 x_2$, respectively, where $\kappa_1$ and $\kappa_2$ are *positive*. Since $\kappa_1$ and $\kappa_2$ are much smaller than $K_1$ and $K_2$, we ignore $\rho_1$ and $\rho_2$.

**B. EM force fluctuations**

We use the same notation as in Part A. Insofar as the gradient force is concerned, the presence of the particle slightly changes the EM energy stored around the focal point of the lens, but does not change the number of the photons passing through the focal plane of the lens. Therefore, the quantum operator corresponding to $g_i$ reads $-A_i P_L x_i$, where $\vec{r} = (x_1, x_2, x_3)$ now denotes the operator corresponding to the position of the particle center, and $P_L$ now denotes the operator corresponding to the optical power carried by the Gaussian beam (and passing through the focal plane of the lens). The Hamiltonian $H_{g_i} = A_i P_L x_i^2 / 2$ commutes with $P_L$.

The relation $\rho_3 = BP_L$ derived in Part A must now be rewritten as $\overline{\rho}_3 = B\overline{P}_L$, where $\overline{O} = \mathrm{E}[O]$ denotes the expectation value of $O$. One might infer from $\overline{\rho}_3 = B\overline{P}_L$ that the operator corresponding to $\rho_3$ reads $BP_L$. Such an inference is incorrect. Here, we rigorously derive the spectral density of $\rho_3$ (see [7] for the definition of spectral density). The approach we adopt yields not only the spectral density of $\rho_3$, but also the expectation value of $\rho_3$ (viz., $\overline{\rho}_3 = B\overline{P}_L$, which was derived in Part A) as well as the spectral densities of the components of the recoil force ($\vec{\sigma}$), whose expectation value is zero (viz., $\mathrm{E}[\vec{\sigma}] = \vec{0}$). The results to be derived here are applicable not only to the laser light trapping the particle (and illuminating it for photodetection), but also to the laser light cooling the particle's motion (in the realizations of feedback cooling which employ laser light to cool the particle's motion).

Assuming that the number of the photons emitted by the laser in any time interval of length $\tau$ has a Poisson distribution with the expectation value $\bar{P}_L \tau / (\hbar \omega_0)$, and that the emission times are independent of each other, the spectral density of $P_L$ is the constant function $\hbar \omega_0 \bar{P}_L$. It is noteworthy that the mean laser power ($\bar{P}_c$) used to cool the particle's motion (in the realizations of feedback cooling which employ laser light to cool the particle's motion) is much smaller than $\bar{P}_L$. Also, unlike the fluctuations $P_L - \bar{P}_L$, which are mainly due to inherent uncertainty in the emission of photons by the laser, the fluctuations $P_c - \bar{P}_c$ are mainly due to fluctuations in the electric current applied to the laser.

We now derive the spectral density of $\rho_3$ for the laser light trapping the particle (and illuminating it for photodetection). The radiation pressure $\hat{z}\rho_3$ comes from the *initial* linear momentum of the photons interacting with the particle. The photons interacting with the particle are either scattered or absorbed by the particle. We write $\rho_3$ as the sum of $\hbar k_z \sum_{m=1}^{N'} \delta(t - t'_m)$ and $\hbar k_z \sum_{m=1}^{N''} \delta(t - t''_m)$, where $\hbar k_z$ is the initial linear momentum of the photons interacting with the particle, the observable $N'$ (or $N''$) is the number of the photons scattered (or absorbed) by the particle in the time interval $(0, T \to \infty)$, and the observables $t'_1, t'_2, ..., t'_{N'}$ (or $t''_1, t''_2, ..., t''_{N''}$) are the times at which the photons are scattered (or absorbed). Since the particle is around the focal point of the lens (viz., $\mathrm{E}[|\vec{r}|] \ll \lambda_0$), Eq. (A1) indicates that $k_z$ is equal to $k_0 - 1/z_0 = k_0(1 - 0.5NA^2)$. Since the number of the photons emitted by the laser in any time interval of length $\tau$ has a Poisson distribution with the expectation value $\bar{P}_L \tau / (\hbar \omega_0)$, and since the emission times are independent of each other, we can say that: (i) $N'$ and $N''$, which are independent of each other, have Poisson distributions with the expectation values $\bar{P}_s T / (\hbar \omega_0)$ and $\bar{P}_a T / (\hbar \omega_0)$, respectively, where $\bar{P}_s = \mathrm{E}[P_s]$ and $\bar{P}_a = \mathrm{E}[P_a]$ are the mean optical powers scattered and absorbed by the particle, respectively, (ii) for a given $N'$, the observables $t'_m$ and $t'_n$ (for $n \neq m$) are independent of each other, (iii) for a given $N''$, the observables $t''_m$ and $t''_n$ (for $n \neq m$) are independent of each other, (iv) for a given $N'$ and $N''$, the observables $t'_m$ and $t''_n$ are independent of each other, and (v) for a given $N'$ and $N''$, the observables $t'_m$ and $t''_m$ each have a uniform distribution over the interval $(0,T)$. Therefore, the expectation value of $\rho_3(t)\rho_3(t+\tau)$ can be written as

$$\mathrm{E}[\rho_3(t)\rho_3(t+\tau)] = \mathrm{E}[\mathrm{E}[\rho_3(t)\rho_3(t+\tau) | N', N'']] =$$

$$\hbar^2 k_z^2 \cdot \mathrm{E}\left[ N' \int_0^T \delta(t-t')\delta(t-t'+\tau) \frac{dt'}{T} \right] + \hbar^2 k_z^2 \cdot \mathrm{E}\left[ (N'^2 - N') \int_0^T \int_0^T \delta(t-t')\delta(t-\xi'+\tau) \frac{dt'd\xi'}{T^2} \right] +$$

$$\hbar^2 k_z^2 \cdot \mathrm{E}\left[ N'' \int_0^T \delta(t-t'')\delta(t-t''+\tau) \frac{dt''}{T} \right] + \hbar^2 k_z^2 \cdot \mathrm{E}\left[ (N''^2 - N'') \int_0^T \int_0^T \delta(t-t'')\delta(t-\xi''+\tau) \frac{dt''d\xi''}{T^2} \right] +$$

$$\hbar^2 k_z^2 \cdot \mathrm{E}\left[ N'N'' \int_0^T \int_0^T \delta(t-t')\delta(t-t''+\tau) \frac{dt'dt''}{T^2} \right] + \hbar^2 k_z^2 \cdot \mathrm{E}\left[ N''N' \int_0^T \int_0^T \delta(t-t'')\delta(t-t'+\tau) \frac{dt''dt'}{T^2} \right] =$$

$$\frac{\hbar k_z^2 \bar{P}_s}{\omega_0} \delta(\tau) + \frac{k_z^2 \bar{P}_s^2}{\omega_0^2} + \frac{\hbar k_z^2 \bar{P}_a}{\omega_0} \delta(\tau) + \frac{k_z^2 \bar{P}_a^2}{\omega_0^2} + \frac{k_z^2 \bar{P}_s \bar{P}_a}{\omega_0^2} + \frac{k_z^2 \bar{P}_a \bar{P}_s}{\omega_0^2} =$$

$$\frac{\hbar k_z^2 (\bar{P}_s + \bar{P}_a)}{\omega_0} \delta(\tau) + \frac{k_z^2 (\bar{P}_s + \bar{P}_a)^2}{\omega_0^2}. \tag{B1}$$

The square root of the second term on the right hand side of Eq. (B1) is the expectation value of $\rho_3$. Moreover, the Fourier transform of the first term on the right hand side of Eq. (B1) with respect to $\tau$ is the spectral density of $\rho_3$.

We now derive $\bar{P}_s$ and $\bar{P}_a$. Under the dipole approximation, the electric field radiated by the particle ($\vec{E}_r$) is equal to the electric field radiated by a point-like dipole with the electric dipole moment $\hat{x}\,\text{Re}[e^{-i\omega_0 t}\alpha E_{inc}(\vec{r})]$, where $\alpha$ denotes the polarizability of the particle, which was given in Part A, and $E_{inc}(\vec{r})$ is the function given by Eq. (A1) and evaluated at $\vec{r}=(x_1,x_2,x_3)$ [2,3]. Since the particle is around the focal point of the lens (viz., $\text{E}[|\vec{r}|]\ll \lambda_0$), its dipole moment can be approximated by $\hat{x}\,\text{Re}[e^{-i\omega_0 t}\alpha E_0 e^{ik_z x_3}]$, where $k_z$ reads $k_0 - 1/z_0$. For a given $\vec{r}$, $\text{E}\left[\vec{E}_r \times \vec{H}_r\right]$ in the far-field is found to be $|\alpha|^2 NA^2 k_0^6 \bar{P}_L \sin^2(\theta')\hat{r}'/(32\pi^3 \varepsilon_0^2 r'^2)$, where $\text{E}\left[E_0^\dagger E_0\right]$ has been replaced by $4\eta_0 \bar{P}_L/(\pi w_0^2)$, and $(r',\theta',\varphi')$ denotes the position of the observation point with respect to the particle center in the spherical coordinate system whose zenith direction is parallel to the $x$ axis. Therefore, $\bar{P}_s$, which is equal to the flux of $\text{E}\left[\vec{E}_r \times \vec{H}_r\right]$ over any closed surface enclosing the particle, reads $|\alpha|^2 NA^2 k_0^6 \bar{P}_L/(12\pi^2 \varepsilon_0^2)$. Also, $\bar{P}_a$ is equal to $(\alpha_I \omega_0/2)\text{E}\left[E_0^\dagger E_0\right]-\bar{P}_s$, where $\alpha_I$ denotes the imaginary part of $\alpha$. Therefore, if the particle has low loss [viz., when $\varepsilon_I \ll \varepsilon_R$], $\bar{P}_s$ and $\bar{P}_a$ can be approximated by $4a^2 NA^2 k_0^6 R^6 \bar{P}_L/3$ and $6\varepsilon_I NA^2 k_0^3 R^3 \bar{P}_L/(\varepsilon_R+2)^2$, respectively, where $\varepsilon_R + i\varepsilon_I$ is the relative permittivity of the particle, and $a$ denotes $(\varepsilon_R-1)/(\varepsilon_R+2)$. In such a case, $\bar{P}_a$ is much smaller than $\bar{P}_s$, and can be ignored.

Interestingly, the square root of the second term on the right hand side of Eq. (B1) is in agreement with $\bar{\rho}_3 = B\bar{P}_L = 4a^2(1-0.5NA^2)NA^2 k_0^6 R^6 \bar{P}_L/(3c)$, which was derived in Part A by applying the dipole approximation to the Maxwell stress tensor. Moreover, the spectral density of $\rho_3$ [viz., the Fourier transform of the first term on the right hand side of Eq. (B1) with respect to $\tau$], can be written as $B'^2 S_{P_L}(\omega)$, where $S_{P_L}(\omega) = \hbar \omega_0 \overline{P}_L$ is the spectral density of $P_L$, and $B'$ is equal to $2a(1-0.5NA^2)NAk_0^3R^3/(\sqrt{3}c)$. The expectation value and spectral density of $\rho_3$ are independent of the position of the particle in that the phase $k_z x_3$ in the expression of the dipole moment does not appear in $\vec{E}_r \times \vec{H}_r$. However, it is noteworthy that the phase $k_z x_3$ will be important in deriving the spectral density of the measurement noise in Part C.

The coefficients $B'$ and $B$, which we derived above for the laser light trapping the particle (and illuminating it for photodetection), are not equal. However, this is not the case for the laser light cooling the particle's motion (in the realizations of feedback cooling which employ laser light to cool the particle's motion). The reason is that, unlike the fluctuations $P_L - \overline{P}_L$, which are mainly due to inherent uncertainty in the emission of photons by the laser, the fluctuations $P_c - \overline{P}_c$ in the laser power used to cool the particle's motion are mainly due to fluctuations $I_c - \overline{I}_c$ in the electric current applied to the laser (it is noteworthy that $I_c - \overline{I}_c$ is the sum of the signal needed for feedback cooling and the measurement noise). In the case of $P_c$, the first term on the right hand side of Eq. (B1) can be ignored while the second must be modified. To modify the second term, we assume that: (i) the proportionality constant between $\overline{P}_c^2$ and $\overline{I}_c^2$ is equal to the proportionality constant between $R_{P_c}(\tau)$ and $R_{I_c}(\tau)$, where $R_O(\tau)$ denotes $\mathrm{E}[O(t)O(t+\tau)] - \overline{O}^2$, (ii) $R_{P_c}(\tau)$ is real, and (iii) the joint probability density function of $t'_m$ and $t'_n$ (for a given $N'$, and for $n \neq m$) over the interval $(0,T)$ is equal to $\mathbb{N}(1/T)^2 [\overline{P}_c^{\,2} + R_{P_c}(t'_m - t'_n)]$, where $\mathbb{N} = 1/[\overline{P}_c^{\,2} + (1/T)\int_0^T R_{P_c}(\tau)d\tau]$ is a normalizing constant.

It is noteworthy that the joint probability density function which led to the second term on the right hand side of Eq. (B1) was $(1/T)^2$. Under the three assumptions made above, the proportionality constant between $\overline{\rho}_{3,c}^{\,2}$ and $\overline{P}_c^{\,2}$ is equal to the proportionality constant between $R_{\rho_{3,c}}(\tau)$ and $R_{P_c}(\tau)$. Moreover, the proportionality constant, which is similar to the expression of $B$, reads $4a^2(1-0.5NA_c^2)NA_c^2 k_{0,c}^6 R^6/(3c)$, where the subscript 'c' for $NA$ and $k_0$ emphasizes that the Gaussian beam cooling the particle's motion is not necessarily the same as the Gaussian beam trapping the particle (and illuminating it for photodetection).

We now derive the spectral densities of the components of the recoil force ($\vec{\sigma}$) for the laser light trapping the particle (and illuminating it for photodetection). The recoil force comes from the *final* linear momentum of the photons scattered by the particle. Unlike the gradient force ($\vec{g}$) and radiation pressure ($\rho_3$), the expectation value of the recoil force ($\vec{\sigma}$) is zero. The spectral densities of the components of $\vec{\sigma}$ are given in [8], but our derivation, which is along the lines of Eq. (B1), allows us to make some important points. We write $\sigma_i(t)$ as $-\sum_{m=1}^{N'} \hbar k'_{m,i} \delta(t - t'_m)$, where the observable $N'$ is the number of the photons scattered by the particle in the time interval $(0, T \to \infty)$, the observables $t'_1, t'_2, ..., t'_{N'}$ are the times at which the photons are scattered, and the observable $k'_{m,i}$ is the *i*th component of the *final* wave vector ($\vec{k}'_m$) of the *m*th photon scattered by the particle. One can write $\vec{k}'_{m,1}$, $\vec{k}'_{m,2}$, and $\vec{k}'_{m,3}$ as $k_0 \cos(\theta'_m)$, $k_0 \sin(\theta'_m)\cos(\varphi'_m)$, and $k_0 \sin(\theta'_m)\sin(\varphi'_m)$, respectively, where the observables $\theta'_m$ and $\varphi'_m$ are the zenith and azimuth angles of $\vec{k}'_m$ in a spherical coordinate system whose zenith direction is parallel to the *x* axis. The observable $k'_{m,i}$ is independent of $t'_n$ (for all $n$), and is independent of $k'_{n,i}$ (for $n \neq m$). The observable $N'$ has a Poisson distribution with the expectation value $\bar{P}_s T / (\hbar \omega_0)$. Also, for a given $N'$, the observables $t'_m$ and $t'_n$ (for $n \neq m$) are independent of each other, and have uniform distributions over the interval $(0,T)$. Therefore, the expectation value of $\sigma_i(t)\sigma_i(t+\tau)$ can be written as

$$E[\sigma_i(t)\sigma_i(t+\tau)] = E\big[E[\sigma_i(t)\sigma_i(t+\tau) | N']\big] =$$

$$\hbar^2 E\big[k'^2_{m,i}\big] \cdot E\left[ N' \int_0^T \delta(t-t')\delta(t-t'+\tau) \frac{dt'}{T} \right] + \hbar^2 E^2\big[k'_{m,i}\big] \cdot E\left[ (N'^2 - N') \int_0^T \int_0^T \delta(t-t')\delta(t-\xi'+\tau) \frac{dt'd\xi'}{T^2} \right] =$$

$$\frac{\hbar E\big[k'^2_{m,i}\big] \bar{P}_s}{\omega_0} \delta(\tau) + \frac{E^2\big[k'_{m,i}\big] \bar{P}_s^{\,2}}{\omega_0^2}. \qquad (B2)$$

Since $E\big[k'_{m,i}\big]$ is zero, the second term on the right side of Eq. (B2) is zero, and therefore, the expectation value of $\sigma_i$ is zero. To calculate the spectral density of $\sigma_i$ [viz., the Fourier transform of the first term on the right hand side of Eq. (B2) with respect to $\tau$], $E\big[k'^2_{m,i}\big]$ must be found. The spatial variation $\sin^2(\theta')$ of $E\big[\vec{E}_r \times \vec{H}_r\big]$, which was given below Eq. (B1), means that the joint probably density function of $\theta'_m$ and $\varphi'_m$ in the expression of $\vec{k}'_m$ reads $3\sin^3(\theta'_m)/(8\pi)$, and, as a result, $E\big[k'^2_{m,i}\big]$ is equal to $k_0^2/5$ for $i$=1, and to $2k_0^2/5$ for $i$=2,3. Therefore, the spectral density of $\sigma_i$ reads $C_i^2 \hbar \omega_0 \bar{P}_L$, where $C_i$ is equal to $2aNAk_0^3 R^3/(\sqrt{15}c)$ for $i$=1, and to $2\sqrt{2}aNAk_0^3 R^3/(\sqrt{15}c)$ for $i$=2,3.

Since the second term on the right hand side of Eq. (B2) is independent of $\tau$, it might seem that the fact $E\big[k'_{m,i}\big] = 0$ is always unimportant in deriving the spectral density of $\sigma_i$.

However, in the case of the laser light cooling the particle's motion (in the realizations of feedback cooling which employ laser light to cool the particle's motion), the modified form of the second term on the right hand side of Eq. (B2) is not independent of $\tau$, but the fact $\mathrm{E}\left[k'_{m,i}\right]=0$ ensures that it does not contribute to the spectral density of the recoil force.

### C. Surface temperature and intrinsic damping rate

The gas molecules surrounding the particle exert a damping force $-M\Gamma\vec{v}$ on it, where $M$ and $\vec{v}$ denote the mass and velocity of the particle, respectively. According to Epstein's seminal paper [9], the intrinsic damping rate $\Gamma$ for the motion of a spherical particle of radius $R$ in a rarefied gas can be written as the sum of $\Gamma_{im} = 4\pi\rho_a \bar{v}_{im} R^2/(3M)$, which is the contribution of the gas molecules impinging on the particle, and $\Gamma_{em} = \pi^2 \rho_a \bar{v}_{em} R^2/(6M)$, which is the contribution of the gas molecules emerging from the particle, where $\rho_a = m_a P_{am}/(k_B T_{am})$ is the gas density, $\bar{v}_{im}$ and $\bar{v}_{em}$ read $\sqrt{8k_B T_{am}/(\pi m)}$ and $\sqrt{8k_B T_{em}/(\pi m)}$, respectively, $m$ denotes the mass of the gas molecules, $P_{am}$ denotes the ambient pressure, $T_{am}$ denotes the ambient temperature, and $T_{em}$ denotes the temperature of the gas molecules emerging from the particle. Einstein's formula is applicable whenever $T_{em}$ is definable.

The gas molecules also exert a random force $(f_1, f_2, f_3)$ on the particle. When $T_{em}$ is equal to $T_{am}$, the spectral density of $f_i$ can be derived by using the Caldeira-Leggett model, and can be written as a function $G(\omega; T_{am}, \Gamma)$ of $T_{am}$ and $\Gamma$ [10]. The function $G(\omega; T_{am}, \Gamma)$ can be rewritten as the sum of $G(\omega; T_{am}, \Gamma_{im})$ and $G(\omega; T_{am}, \Gamma_{em})$. When $T_{em}$ is not equal to $T_{am}$, the spectral density can be written as the sum of $G(\omega; T_{am}, \Gamma_{im})$ and $G(\omega; T_{em}, \Gamma_{em})$, and be simplified to $G(\omega;T,\Gamma)$, where $T$ is equal to $(\Gamma_{im}T_{am}+\Gamma_{em}T_{em})/\Gamma$ [11]. The temperature $T$ is what appears in the expressions of $\bar{n}_{th,i}$ and $\bar{n}_i$ (for all $i$) in the main text.

We now derive $T_{em}$. It can be written as $T_{am}+\alpha_{acc}(T_s-T_{am})$, where $\alpha_{acc}$ and $T_s$ denote the thermal accommodation coefficient and surface temperature of the particle, respectively [12,13]. The surface temperature ($T_s$) is the solution to $\bar{P}_a = P_{c.c.} + P_{r.c.}$, where $\bar{P}_a$ denotes the mean optical power absorbed by the particle, and $P_{c.c.}$ and $P_{r.c.}$ denote the rates of heat conduction and thermal radiation from the particle, respectively. The expression of $\bar{P}_a$, which was derived in Part B, reads $6\varepsilon_I NA^2 k_0^3 R^3 \bar{P}_L/(\varepsilon_R+2)^2$. The rate of heat conduction ($P_{c.c.}$) reads $\frac{(\gamma_a+1)P_{am}\bar{v}_{im}}{8(\gamma_a-1)T_{am}}\alpha_{acc}a_P(T_s-T_{am})$, where $a_P=4\pi R^2$ is the surface area of the particle, and $\gamma_a$ denotes the heat capacity ratio of the gas [12,13]. The rate of thermal radiation ($P_{r.c.}$) reads $\sigma_0 a_P(T_s^4-T_{am}^4)$, where $\sigma_0$ denotes the Stefan-Boltzmann constant, and we have assumed that the emissivity of the particle is unity.

**D. Derivation of $S_{n_1}$, $S_{n_2}$, and $S_{n_3}$**

As is usually the case in experiments [8,14-16], we assume that the Gaussian beam trapping the particle illuminates it for photodetection as well. We use the same notations as in Parts A and B. The measurement of the position of the particle center is carried out by the measurement of the EM field intensity [8,14-16]. The electric field is the sum of the incident electric field ($\vec{E}_L$) given by Eq. (A1) and the electric field radiated by the particle ($\vec{E}_r$). Under the dipole approximation, $\vec{E}_r$ is equal to the electric field radiated by a point-like dipole with the electric dipole moment $\hat{x}\mathrm{Re}[e^{-i\omega_0 t}\alpha E_{inc}(\vec{r})]$, where $\alpha$ denotes the polarizability of the particle, and $E_{inc}(\vec{r})$ is the function given by Eq. (A1) and evaluated at the particle center ($\vec{r}$) [2,3]. Since the particle is around the focal point of the lens (viz., $\mathrm{E}[|\vec{r}|] \ll \lambda_0$), its dipole moment can be approximated by $\hat{x}\mathrm{Re}[e^{-i\omega_0 t}\alpha E_0 e^{ik_z x_3}]$, where $\vec{r}$ has been written as $(x_1, x_2, x_3)$, and $k_z$ reads $k_0 - 1/z_0$. It is noteworthy that the phase $k_z x_3$ in the expression of the dipole moment was not important in deriving the expectation value of $\rho_3$ and the spectral densities of $\rho_3$ and $\sigma_i$, but it is now important in deriving $S_{n_3}$ (which is defined in the main text).

At an observation point $(X, Y, Z)$ far enough from the particle and close enough to the axis of the beam [viz., $Z > 10\lambda_0$ and $X^2, Y^2 < \lambda_0 Z/(20\pi)$], the sum of $\vec{E}_L$ and $\vec{E}_r$ can be written as

$$\vec{E}_d = \hat{x}\mathrm{Re}\left\{e^{-i\omega_0 t}(-i)E_0[z_0/Z + ak_0^3 R^3 X x_1/Z^2 + ak_0^3 R^3 Y x_2/Z^2 + ak_0^2 R^3 x_3/(z_0 Z)]\right\}, \qquad (D1)$$

where $a$ and $R$ denote $(\varepsilon_R - 1)/(\varepsilon_R + 2)$ and the radius of the particle, respectively, and $z_0$ denotes the Rayleigh range of the Gaussian beam. It is emphasized that $(X, Y, Z)$ and $\vec{r} = (x_1, x_2, x_3)$ have been defined with respect to the focal point of the lens employed to generate the Gaussian beam illuminating (and trapping) the particle. In fact, Eq. (D1) is the electric field calculated within the dipole approximation, the far-field approximation, the paraxial approximation, and the assumption that the particle is around the focal point of the lens. The electric field *operator* at $(X, Y, Z)$ is also given by Eq. (D1) if $\vec{r} = (x_1, x_2, x_3)$ and $E_0$ are interpreted as operators. The operator corresponding to the optical power carried by the Gaussian beam can be written as $P_L = \pi w_0^2 E_0^\dagger E_0/(4\eta_0)$ in terms of the operator $E_0$, where $w_0$ denotes the minimum beam radius of the Gaussian beam, and $\eta_0$ denotes the impedance of free space.

The spectral density of the photocurrent ($I_i$) generated by a small enough photodetector centered at ($X_i, Y_i, Z_i$) can be written as the sum of $S_{M_i}(\omega)$ and $S_{N_i}(\omega)$, where $S_{N_i}(\omega)$, which is independent of $\omega$ over the detection bandwidth, reads $\zeta_i q^2 \mathrm{E}\left[ E_{d_i}^\dagger E_{d_i} \right]$, and $S_{M_i}(\omega)$ is equal to the Fourier transform of

$$R_{M_i}(\tau) = \zeta_i^2 q^2 \mathrm{E}\left[ E_{d_i}^\dagger(t) E_{d_i}^\dagger(t+\tau) E_{d_i}(t+\tau) E_{d_i}(t) \right] - \zeta_i^2 q^2 \mathrm{E}\left[ E_{d_i}^\dagger E_{d_i}^\dagger E_{d_i} E_{d_i} \right] \tag{D2}$$

with respect to $\tau$ [17]. The role of the second term on the right hand side of Eq. (D2) is to eliminate the $\tau$-independent part of the first term. The coefficient $q$ denotes the elementary charge, and the coefficient $\zeta_i$ reads $a_{d_i}/(2\eta_0 \hbar \omega_0)$ in terms of the area of the photodetector ($a_{d_i}$). We have assumed that the detection efficiency is unity, and the detection bandwidth is large in comparison with $\Omega_i$. The subscript $i$ emphasizes that the photodetector is intended to measure $x_i$. We can interpret the fluctuations $I_i - \overline{I}_i$ as an incoherent sum of the signal $M_i$ and the measurement noise $N_i$. We write $I_i - \overline{I}_i$ as $M_i \oplus N_i$.

Given Eq. (D1) and the fact that $\mathrm{E}[|\vec{r}|]$ is much smaller than $\lambda_0$, $S_{N_i}(\omega)$, which is independent of $\omega$, is found to be $\zeta_i q^2 \mathrm{E}\left[ E_0^\dagger E_0 \right](z_0/Z_i)^2$. We now derive $S_{M_i}(\omega)$ for each $i$.

The photodetector intended to measure $x_3$ is centered at $(0,0,Z)$, which allows us to ignore any signature of $x_1$ and $x_2$ in $M_3$. Ignoring $(ak_0^2 R^3/(z_0 Z))^4 \mathrm{E}\left[ x_3(t) x_3^2(t+\tau) x_3(t) - x_3^4 \right]$, we find that $S_{M_3}(\omega)$ can be written as $\theta_3 S_{x_3}(\omega)$, where the coefficient $\theta_3$ reads $(2\zeta_3 q a k_0^2 R^3/Z^2)^2 \mathrm{E}\left[ E_0^{\dagger 2} E_0^2 \right]$. Therefore, $S_{n_3}$, defined in the main text, is found to be

$$S_{n_3} = S_{N_3}/\theta_3 = \frac{\pi \hbar \omega_0 w_0^2 z_0^2}{8a^2 k_0^4 R^6 \overline{P}_L} \frac{Z^2}{a_{d_3}}. \tag{D3}$$

The expression of $S_{n_3}$ indicates that it is advantageous to increase $a_{d_3}$, but the condition of the paraxial approximation must not be violated. The maximum allowable value of $a_{d_3}$ is around $\lambda_0 Z/(5\pi)$ at a given $Z$. Assuming that $a_{d_3} \approx \lambda_0 Z/(5\pi)$, $S_{n_3}$ is proportional to $Z$. It is emphasized that the far-field approximation requires $Z$ to be kept well above $\lambda_0$ (viz., $Z > 10\lambda_0$).

To measure $x_1$, the photocurrents $I_1 = \overline{I}_1 + M_1 \oplus N_1$ and $I'_1 = \overline{I}'_1 + M'_1 \oplus N'_1$ generated by two photodetectors centered at $(X,0,Z)$ and $(-X,0,Z)$, respectively, are subtracted from each other [16]. Such a balanced detection allows us to ignore any signature of $x_3$ (and $x_2$) in $I_1 - I'_1$. It should be noted that $I_1 - I'_1$ cannot be written as $(\overline{I}_1 - \overline{I}'_1) + (M_1 - M'_1) \oplus (N_1 - N'_1)$. In fact, $M_1$ and $-M'_1$ are added coherently (viz., they are perfectly correlated) while $N_1$ and $-N'_1$ are added incoherently (viz., they are uncorrelated). Therefore, we write $I_1 - I'_1$ as $(\overline{I}_1 - \overline{I}'_1) + (M_1 - M'_1) \oplus [N_1 \oplus (-N'_1)]$. The spectral density of $N_1 \oplus (-N'_1)$ is equal to $2S_{N_1}(\omega)$ while the spectral density of $(M_1 - M'_1)$ can be written as $4\theta_1 S_{x_1}(\omega)$. Ignoring $(ak_0^3 R^3 X/Z^2)^4 \mathrm{E}\left[x_1(t)x_1^2(t+\tau)x_1(t) - x_1^4\right]$, we find that $\theta_1$ reads $(2\zeta_1 q a k_0^3 R^3 z_0 X/Z^3)^2 \mathrm{E}\left[E_0^{\dagger 2} E_0^2\right]$. Therefore, $S_{n_1}$, defined in the main text, reads

$$S_{n_1} = 0.5\, S_{N_1}/\theta_1 = 0.5 \frac{\pi \hbar \omega_0 w_0^2}{8a^2 k_0^6 R^6 \overline{P}_L} \frac{Z^4}{X^2 a_{d_1}}. \tag{D4}$$

The expression of $S_{n_1}$ indicates that it is advantageous to increase $X^2 a_{d_1}$, but the condition of the paraxial approximation must not be violated. The maximum allowable value of $X^2 a_{d_1}$ is around $[\lambda_0 Z/(45\pi)]^2$ at a given $Z$. It happens when $X^2 \approx a_{d_1} \approx \lambda_0 Z/(45\pi)$. Assuming that $X^2 \approx a_{d_1} \approx \lambda_0 Z/(45\pi)$, $S_{n_1}$ is proportional to $Z^2$. It is emphasized that the far-field approximation requires $Z$ to be kept well above $\lambda_0$ (viz., $Z > 10\lambda_0$).

To measure $x_2$, the photocurrents generated by two photodetectors centered at $(0, Y, Z)$ and $(0, -Y, Z)$ are subtracted from each other. The term $S_{n_2}$, defined in the main text, is derived in the same way as $S_{n_1}$ was derived. It reads

$$S_{n_2} = 0.5\, S_{N_2}/\theta_2 = 0.5 \frac{\pi \hbar \omega_0 w_0^2}{8 a^2 k_0^6 R^6 \bar{P}_L} \frac{Z^4}{Y^2 a_{d_2}}. \tag{D5}$$

The maximum allowable value of $Y^2 a_{d_2}$ is around $[\lambda_0 Z/(45\pi)]^2$ at a given $Z$. It happens when $Y^2 \approx a_{d_2} \approx \lambda_0 Z/(45\pi)$. Assuming that $Y^2 \approx a_{d_2} \approx \lambda_0 Z/(45\pi)$, $S_{n_2}$ is proportional to $Z^2$.

It should be noted that we have approximated $\mathrm{E}[x_i(t) x_i(t+\tau)]$ by its real part when deriving $S_{M_i}(\omega)$. In other words, we have approximated $\mathrm{E}\left[E_{d_i}^\dagger(t) E_{d_i}^\dagger(t+\tau) E_{d_i}(t+\tau) E_{d_i}(t)\right]$ by $\mathrm{E}\left[E_{d_i}^\dagger(t) E_{d_i}(t) E_{d_i}^\dagger(t+\tau) E_{d_i}(t+\tau)\right]$, which is proportional to $\mathrm{E}\left[P_{d_i}(t) P_{d_i}(t+\tau)\right]$, where $P_{d_i}$ denotes the optical power received by the photodetector. This approximation was necessary in deriving the relation $S_{M_i}(\omega) = \theta_i S_{x_i}(\omega)$.

Finally, it should be noted that in practice the photodetectors cannot be placed at a small distance of $10\lambda_0$ from the particle, but rather at a distance of $10^6 \lambda_0 - 10^7 \lambda_0$ from it. The effect of such a long distance on $S_{n_i}$ can be compensated to some extent by employing a collimating lens with a large enough numerical aperture to collect a large enough amount of light before directing the light to the photodetectors. The focal point of the collimating lens coincides with the focal point of the lens employed to generate the Gaussian beam illuminating (and trapping) the particle. Also, each photodetector is placed at the focal plane of a converging lens which compensates the Fourier transforming effect of the collimating lens. The results derived above all remain valid, but $Z$ no longer denotes the *actual* distance between the photodetector and the focal point of the lens employed to generate the Gaussian beam illuminating (and trapping) the particle. Also, $a_{d_i}$ no longer denotes the *actual* area of the photodetector. Rather, $Z$ is a parameter determined by the characteristics of the collimating lens, the converging lens, and other optical devices between them (viz., beam splitters and mirrors), hence the name '*effective distance*' in the main text. The parameters $Z$ must still meet the condition $Z > 10\lambda_0$. Also, the maximum allowable values of the parameters $a_{d_3}$, $X^2 a_{d_1}$, and $Y^2 a_{d_2}$ are still $\lambda_0 Z/(5\pi)$, $[\lambda_0 Z/(45\pi)]^2$, and $[\lambda_0 Z/(45\pi)]^2$, respectively.

* jazayeri@ee.sharif.edu

1. B. E. A. Saleh and M. C. Teich, *Fundamentals of Photonics* (John Wiley & Sons, New York, 1991).
2. C. F. Bohren and D. R. Huffman, *Absorption and Scattering of Light by Small Particles* (John Wiley & Sons, New York, 1998).
3. B. T. Draine, Astrophys. J. 333, 848 (1988).
4. A. M. Jazayeri and K. Mehrany, Phys. Rev. A 89, 043845 (2014).
5. P. C. Chaumet and A. Rahmani, Opt. Exp. 17, 2224 (2009).
6. A. M. Jazayeri and K. Mehrany, J. Opt. Soc. Am. B 34, 2179 (2017).
7. We denote the spectral density of an observable $O$ by $S_O(\omega)$, and define it as the Fourier transform of $R_O(\tau) = \mathrm{E}[O(t)O(t+\tau)] - \mathrm{E}^2[O(t)]$ with respect to $\tau$, provided that the expectation values, denoted by $\mathrm{E}[\cdot]$, are independent of *t*. The spectral density is a real function, and sometimes an even function. It is always an even function in the classical limit.
8. V. Jain, J. Gieseler, C. Moritz, C. Dellago, R. Quidant, and L. Novotny, Phys. Rev. Lett. 116, 243601 (2016).
9. P. S. Epstein, Phys. Rev. 23, 710 (1924): Equation (29).
10. A. O. Caldeira and A. J. Leggett, Phys. Rev. Lett. 46, 211 (1981).
11. J. Millen, T. Deesuwan, P. Barker, and J. Anders, Nat. Nanotechnol. 9, 425 (2014).
12. M. Knudsen, Ann. Phys. 339, 593 (1911).
13. E. H. Kennard, *Kinetic Theory of Gases* (McGraw-Hill, New York, 1938).
14. T. Li, S. Kheifets and M. G. Raizen, Nat. Phys. 7, 527 (2011).
15. F. Tebbenjohanns, M. Frimmer, A. Militaru, V. Jain, and L. Novotny, arXiv:1812.09875.
16. J. Gieseler, B. Deutsch, R. Quidant, and L. Novotny, Phys. Rev. Lett. 109, 103603 (2012).
17. H. J. Carmichael, J. Opt. Soc. Am. B 4, 1588 (1987).